\documentclass[%
 reprint,
superscriptaddress,
 amsmath,amssymb,
 aps,
 prl,
]{revtex4-1}

\usepackage[dvips]{graphicx}
\usepackage{dcolumn}
\usepackage{bm}
\usepackage{xcolor}

\begin{document}

\preprint{APS/123-QED}

\title{Viscous fingering and dendritic growth under an elastic membrane}

\author{L. Duclou\'e}
\affiliation{%
 Manchester Centre for Nonlinear Dynamics and School of Physics and Astronomy, University of Manchester, Oxford Road, Manchester M13 9PL, UK}
\author{A. L. Hazel}%
\affiliation{%
 Manchester Centre for Nonlinear Dynamics and School of Mathematics, University of Manchester, Oxford Road, Manchester M13 9PL, UK}
\author{D. Pihler-Puzovi\'c}%
\affiliation{%
 Manchester Centre for Nonlinear Dynamics and School of Physics and Astronomy, University of Manchester, Oxford Road, Manchester M13 9PL, UK}
 \author{A. Juel}%
\affiliation{%
 Manchester Centre for Nonlinear Dynamics and School of Physics and Astronomy, University of Manchester, Oxford Road, Manchester M13 9PL, UK}

\date{\today}

\begin{abstract}
We investigate the viscous fingering instability that arises when air is injected from the end of an oil-filled, compliant channel. We show that induced axial and transverse depth gradients foster novel pattern formation. Moreover, the steady propagation of the interface allows us to elucidate the nonlinear saturation of a fingering pattern first observed in a time-evolving system (Pihler-Puzovi\'c et al. PRL 108, 074502, 2012): the wavelength is set by the viscous fingering mechanism, but the amplitude is inversely proportional to the tangent of the compliant wall's inclination angle.\\

\end{abstract}

\pacs{Valid PACS appear here}
\maketitle
\paragraph*{}
Two-phase displacement flows in narrow gaps are ubiquitous in industrial, geophysical and biological processes. Examples include coating processes~\citep{coyle1986film}, enhanced oil recovery~\citep{sheng2010modern}, $\mathrm{CO}_2$ sequestration~\citep{huppert2014fluid}, soil drying~\citep{bachmann2002review} and lung biomechanics~\citep{heil2011fluid}. 
When a less viscous fluid is injected into a more viscous layer, the interface between the two fluids is linearly unstable and the most unstable wavenumber increases with the capillary number -- the ratio of destabilizing viscous forces to stabilizing surface tension forces~\citep{saffman1958penetration}. This viscous fingering instability has been widely studied in Hele-Shaw cells as an archetype for pattern-forming interfacial instabilities in other contexts such as crystallization or bacterial growth~\citep{arneodo1989uncovering,ben1992adaptive, ben1990formation}. The introduction of depth gradients transverse to the driving direction (grooves and threads~\citep{rabaud1988dynamics}, or step-like constrictions~\citep{franco2016sensitivity}) promotes the development of transverse patterns. In contrast, a convergent depth gradient in the driving direction, introduced in a rigid channel~\citep{al2012control} or through a compliant boundary~\citep{pihler2012suppression}, considerably delays the onset of the axial fingering instability. In a compliant radial Hele-Shaw cell, the interface was observed to destabilize into a ring of short constant-depth fingers~\citep{pihler2012suppression} beyond a critical value of the flow rate, entirely different to the highly-branched patterns formed in rigid cells of uniform depth. In this axisymmetric geometry, however, the radial velocity of the constant-flux driven air-liquid interface decreases continuously, reducing the effective capillary number and thus the growth rate of the linear instability, so that ultimately, the compliant axisymmetric system always re-stabilizes.
\begin{figure}
\centering
\includegraphics[width=1.0\linewidth]{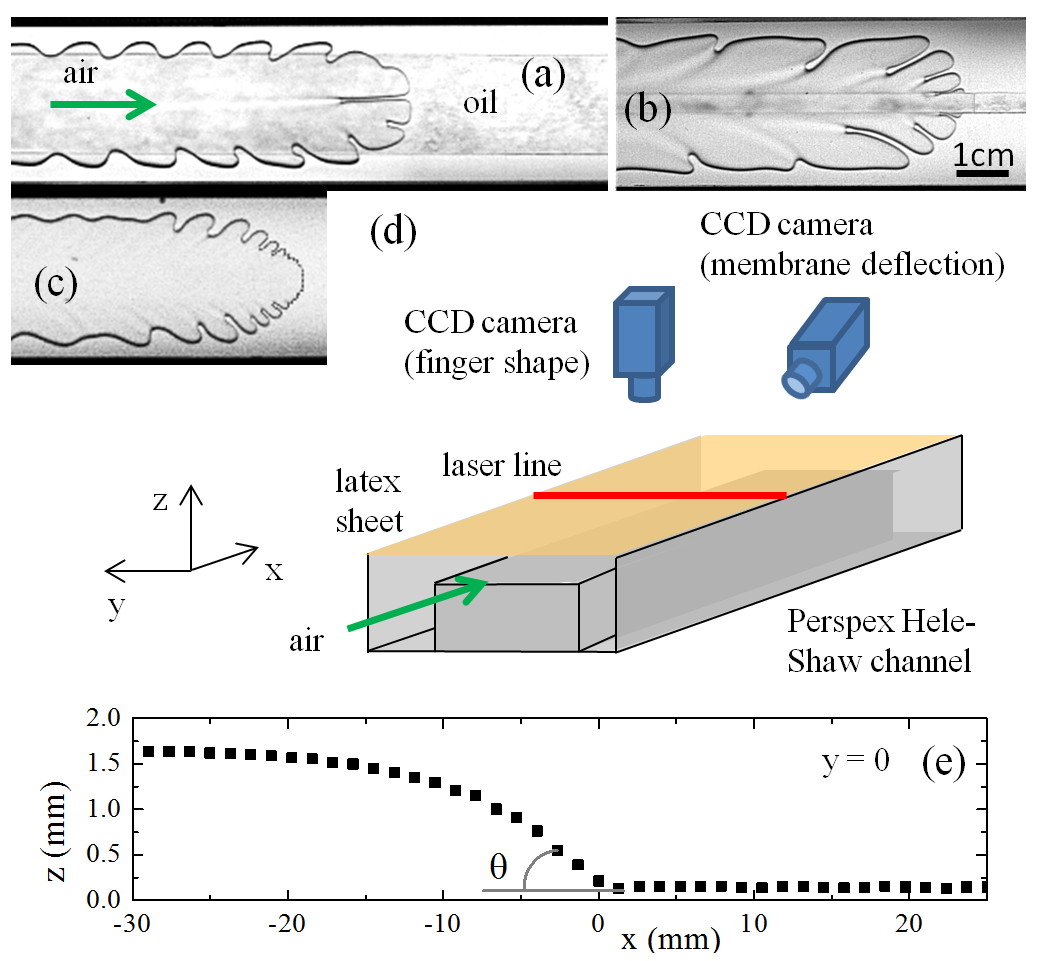}
\caption{ (a), (b), (c): Examples of the patterns that can be promoted through axial and transverse depth gradients in a compliant Hele-Shaw channel. Centred block dimensions (width $\times$ heigth): (a) 20~mm $\times$ 900~$\mu$m, (b) 4~mm $\times$ 380~$\mu$m, (c) 8~mm $\times$ 60~$\mu$m. (d) Experimental set-up, made of a rigid Hele-Shaw channel of non-uniform depth topped with an elastic membrane. Nitrogen is injected at a constant flow rate from one end of the oil-filled channel. (e) Axial profile along the centreline of the channel showing the deflection of the top membrane during the steady propagation of an air finger into oil. The air-oil interface sits at $x=0$ and $z=0$ is taken at the surface of the block. \label{fig:set-up}}
\end{figure}
\paragraph*{}
In this letter, we explore the nonlinear saturation of viscous fingering patterns in a rectangular channel with a deformable top boundary. This geometry circumvents the time-dependence of the radial cell by supporting the steady propagation of an air-oil interface driven by a constant volume flux. However, the deformation of the upper boundary, pinned on both sidewalls of the channel, introduces both axial and transverse depth gradients, which results in novel pattern selection mechanisms. We show that non-uniform depth can not only suppress viscous fingering but also enhance it, and can be used to create and control a wide variety of interfacial patterns (Fig.~\ref{fig:set-up}(a--c)). In order to focus on pattern selection resulting from an axial depth gradient, we choose a regime in which a flat front normal to the driving direction steadily propagates in our deformable channel~\citep{ducloue2016reopening}  (Fig.~\ref{fig:set-up}(d)). This front fingers depending on two parameters: (\textit{i}) the value of the capillary number $\mathrm{Ca}=\mu U/\sigma$ (with $U$ the speed of the interface, $\mu$ the dynamic viscosity of the oil and $\sigma$ its surface tension) and (\textit{ii}) the extensional stiffness~\citep{peng2015displacement} of the flexible boundary quantified by $T=Eb/(12(1-\nu^2))$ (with $E$ the Young's modulus, $b$ the thickness and $\nu$ the Poisson's ratio of the flexible boundary). By varying $T$, we explore the transition of the instability to the classical Saffman-Taylor finger as the system approaches the limit of a rigid Hele-Shaw channel. 
\paragraph*{}
The transverse curvature induced by the deformation of the elastic upper boundary means that the propagation of a flat interface spanning the width of the channel is never observed. Instead, the inflation of the membrane couples with the oil flow at the propagation front to select the shape of an air finger depending on $\mathrm{Ca}$ and the initial cross-sectional shape of the channel set prior to experimentation. Strong initial collapse of the membrane onto the channel floor produces a wide air finger, which exhibits a ``flat tip": the reopening air finger redistributes the thin oil layer in the collapsed region via an elastic peeling mode~\citep{ducloue2016reopening}. Thus, the flat section of interface propagates steadily in a tapered channel of constant taper angle $\theta$ (see axial profile in Fig.~\ref{fig:set-up}(e)).
\paragraph*{}
Our elasto-rigid Hele-Shaw channel is similar to that described in~\cite{ducloue2016reopening}: the 60~cm long, 30~mm wide and $H=1.05\pm 0.01$~mm deep channel was milled in a Perspex block and topped with a latex membrane (Supatex) of Young's modulus $E=1.44\pm 0.05$~$\mathrm{MPa}$~\citep{pihler2012suppression}, Poisson's ratio $\nu=0.5$ and thickness $b$ chosen between $210$~$\mathrm{\mu m}$ and $570$~$\mathrm{\mu m}$. For all membranes, $b$ was uniform to a precision of 10~$\mathrm{\mu m}$. A pre-tension was applied along the width of the membrane by uniformly hanging weights to a free-hanging long edge before clamping all four edges. This procedure resulted in the membrane lying flat on top of the empty channel and being pre-stressed with a tension of $2.0 \pm 0.1$~$\mathrm{kPa}$ for $b=210$~$\mathrm{\mu m}$, $4.1 \pm 0.1$~$\mathrm{kPa}$ for $b=340$~$\mathrm{\mu m}$ and $10.0 \pm 0.2$~$\mathrm{kPa}$ for $b=460$~$\mathrm{\mu m}$. The thickest membrane could not be laid flat by using this procedure and manual pulling resulted in a tension greater than 10 $\mathrm{kPa}$ but of at least one order of magnitude smaller than $E$. We monitored the propagation of the interface using a CCD camera placed above the channel. We recorded the deflection of the latex membrane as the interface propagated by vertically shining a laser line at a fixed position across the channel width and imaging it with a second CCD camera aligned with the channel axis and tilted by a known angle from the horizontal (between 22$^{\circ}$ and 30$^{\circ}$). This procedure allowed us to measure the vertical displacement of the membrane with a precision of $20$~$\mu$m. Prior to each experiment, the channel was filled with silicone oil (Basildon Chemicals) of viscosity $\mu = 0.099$~Pa~s and surface tension $\sigma = 21$~mN/m at the laboratory temperature of 21$^{\circ}$C, which fully wets the latex and the Perspex. The initial shape of the latex membrane prior to experimentation could be set by imposing a constant hydrostatic pressure in the oil (relative to atmospheric pressure)~\cite{ducloue2016reopening}. At the end of the filling procedure, the oil was at rest and the membrane was horizontal in the absence of loading, or inflated (deflated) if fluid had been added (withdrawn) from the channel, respectively.
\begin{figure}
\centering
\includegraphics[width=1.0\linewidth]{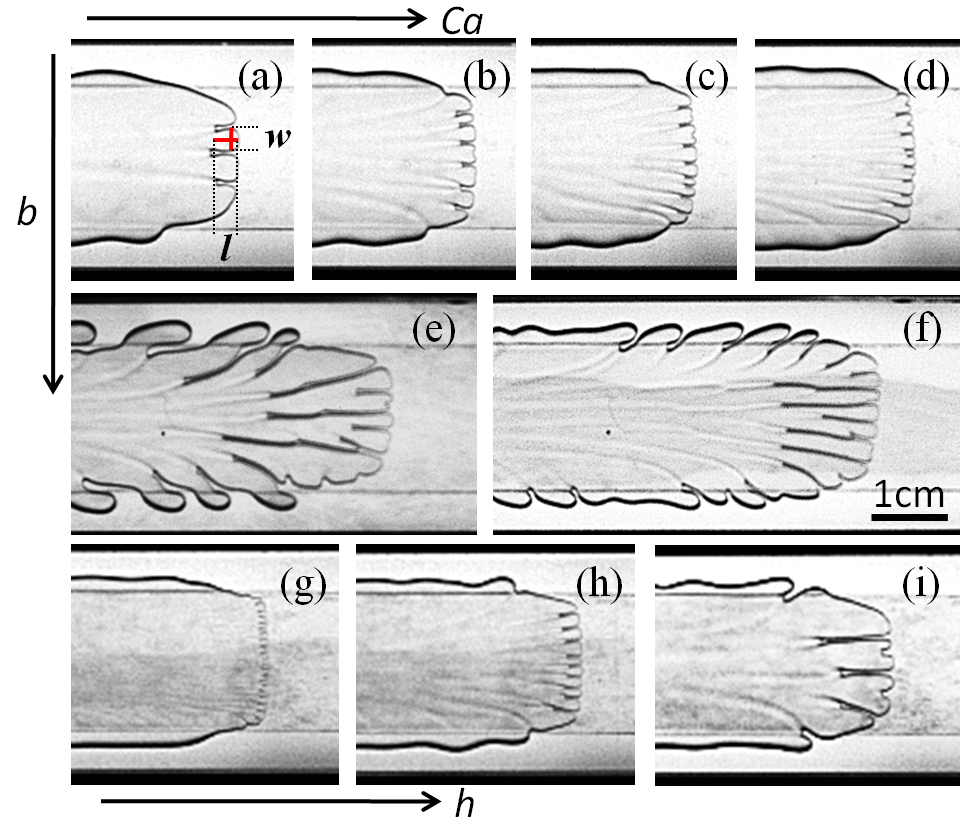}
\caption{Typical images of the axial fingering pattern. First row: $b=210$~$\mu$m, $h=150$~$\mu$m and Ca=0.21 (a), Ca=0.56 (b), Ca=1.05 (c) and Ca=1.74 (d); second row: $b=570$~$\mu$m, $h=150$~$\mu$m and Ca=1.61 (e) and Ca=3.77 (f); third row: $b=340$~$\mu$m, Ca=0.62$\pm$0.02 and $h=100$~$\mu$m (g), $h=130$~$\mu$m (h) and $h=240$~$\mu$m (i).\label{fig:fingering}}
\end{figure}
\paragraph*{}
Instead of collapsing the channel to create the transverse depth gradients necessary to support a flat-fronted steady propagation mode, we introduced  an axially uniform rectangular block bonded centrally along the bottom boundary ($20 \pm 0.5$~mm wide, $900\pm10$~$\mu$m high, which is greater than 85\% of $H$), so that the elastic peeling mode could be supported with an initially undeformed top compliant boundary (Fig. 1(d)). In addition, the depth $h$ of the fluid layer on the block could be changed by collapsing the membrane. The elastic peeling mode required sufficiently large values of $\mathrm{Ca}$ to develop, thus introducing a lower bound, dependent on $T$, on the range of $\mathrm{Ca}$ which could be investigated. For small $T$ reached when using a thin silicone membrane ($b=100$~$\mu$m, from Silex), the peeling front remains flat (see video V1 in Supplementary Material), whereas for the larger values of $T$ obtained with our latex membranes, it develops an array of finite-depth fingers actuated by the axial depth gradient quantified by the angle $\theta$ (Fig.~\ref{fig:set-up}(e)). These are clearly visible in Fig.~\ref{fig:fingering} but also present in Figs.~\ref{fig:set-up} (b) and (c), albeit too small to be resolved in the pictures. Similar constant-depth fingers have been observed following the onset of fingering in a compliant radial Hele-Shaw cell~\citep{pihler2012suppression}, but also in the regular pattern formed by the peeling of viscous adhesives~\citep{mcewan1966peeling} or the destabilisation of the meniscus entrained between two rotating cylinders (printer's instability)~\citep{rabaud1991wavelength}. 
\paragraph*{}
Changes to the height and width of the block, as well as to the level of initial collapse can be applied to modify the depth gradients in the channel, and thus generate a wide variety of fingering patterns. The transverse depth gradient at the edge of the occlusion can make the interface develop spatially periodic sideways fingers (Fig.~\ref{fig:set-up}(a)). These patterns are shed from the propagating front, that undergoes oscillations in width driven by the change in cross-sectional curvature of the interface as it crosses the edge of the block. This mechanism has been described in rigid constricted tubes~\citep{pailha2012oscillatory, thompson2014multiple, franco2016sensitivity} and is also the dominant mechanism in Fig.~\ref{fig:set-up}(b), where fingers are generated in the direction of the strongest depth gradient. Axial and transverse depth gradients can also couple to amplify the axial fingers at the edge of the occlusion, creating the dendritic-like pattern illustrated in Fig.~\ref{fig:set-up}(c). The amplification of the fingers generated at the edge of the occlusion leads to approximate proportionate growth (fixed aspect ratio) of the dendritic pattern, a feature which has been observed in other viscous fingering experiments~\citep{bischofberger2014fingering}. The growth of dendrites has been triggered in rigid Hele-Shaw cells by locally perturbing the interface at the finger tip~\citep{couder1986dendritic}, but the selection of the pattern is unexplained: the well-controlled geometry of the compliant channel could enable a more systematic investigation.
\paragraph*{}
We now focus solely on the fingering of the peeling front, which is driven by axial depth gradients. We investigate pattern selection in terms of the average width $w$ and length $l$ of the fingers (defined in Fig.~\ref{fig:fingering}(a)), as a function of $\mathrm{Ca}$, fluid layer thickness $h$ and membrane thickness $b$. The qualitative evolution of the pattern with Ca ($h=150$~$\mu$m) is presented for the thinnest and thickest membranes we used in Fig.~\ref{fig:fingering}(a--d) and (e--f), respectively: for a given membrane, the fingers become narrower and shorter as Ca increases, and we observe much longer fingers with the thicker membrane. At fixed Ca, the fingers widen and lengthen as $h$ increases as shown by Fig.~\ref{fig:fingering}(g--i). As suggested in Fig.~\ref{fig:fingering}(i) and shown in video V2 of the Supplementary Material, the fingers are never strictly steady: they widen and split regularly, but maintain a constant average width and length. This behavior is the main source of uncertainty in our quantitative measurements. For low Ca, large $h$ or thick membranes, the front becomes increasingly curved, patterns become less regular and large amplitude tip-splitting occurs: the rear meniscus and front tip of the fingers no longer travel at the same speed, so that their length increases as they propagate. We do not consider those cases in our quantitative analysis, but report them with empty symbols in the subsequent graphs.
\begin{figure}
\centering
\includegraphics[width=1.0\linewidth]{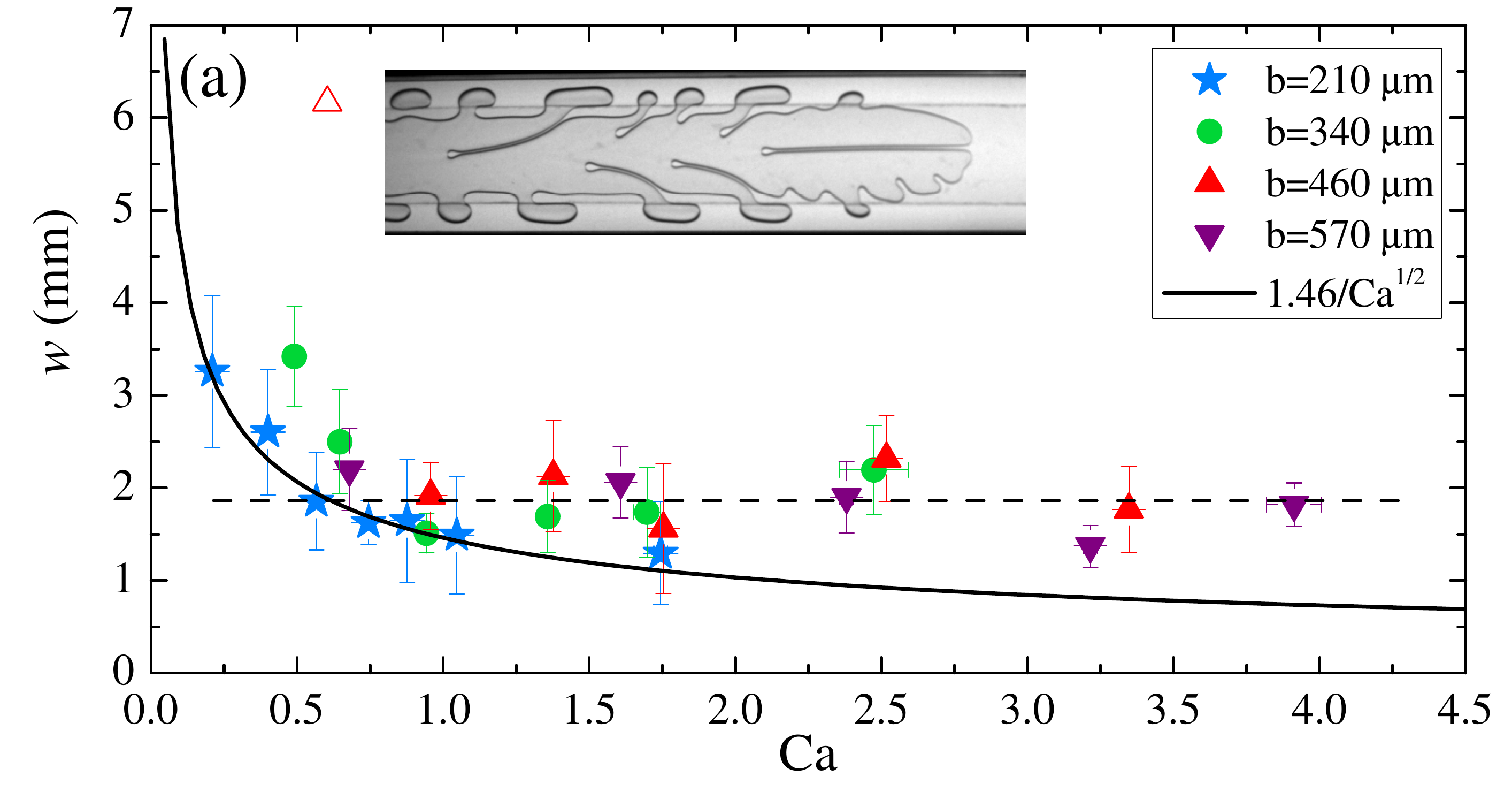}
\includegraphics[width=1.0\linewidth]{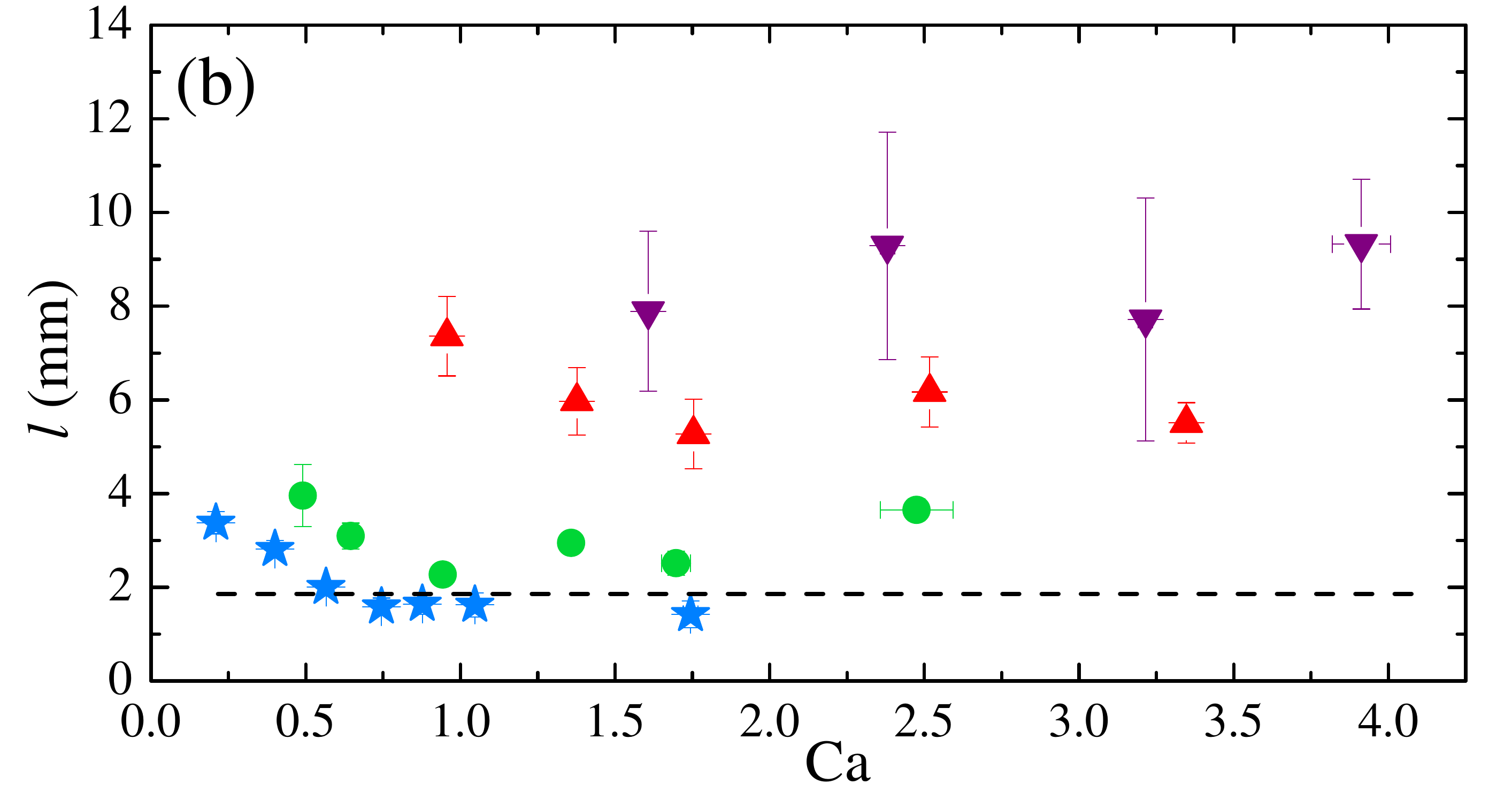}
\caption{Evolution of the finger width (a) and length (b) with Ca for the 4 different latex membranes used ($h=150$~$\mu$m). The solid line is a least-squares fit to describe the $1/\sqrt{Ca}$ behavior of $w$ at low Ca. The dotted guideline present on both figures shows the saturation value of $w$. The empty symbol represents a large amplitude tip-splitting mode, illustrated in the inset picture.\label{fig:geom_of_Ca}}
\end{figure}
\paragraph*{}
The width and length of the fingers are quantified in Fig.~\ref{fig:geom_of_Ca}  as a function of $\mathrm{Ca}$, for $h=150$~$\mu$m. For all values of the membrane thickness, the average finger width decreases similarly with $\mathrm{Ca}$, towards a constant value which does not depend on membrane thickness to within experimental resolution (Fig.~\ref{fig:geom_of_Ca}(a)). For each membrane thickness, the average length of the fingers also decreases towards an approximately constant value with Ca, but this saturation value increases considerably with membrane thickness. Moreover, the value of $\mathrm{Ca}$ at which saturation is reached increases with membrane thickness. For the thickest membrane only the saturated regime could be explored because the elastic peeling front only formed at large $\mathrm{Ca}$ (smaller $\mathrm{Ca}$ resulted in large amplitude tip-splitting of a curved front  pictured in the inset in Fig.~\ref{fig:geom_of_Ca}(a)). For a given Ca chosen in the saturated plateau, the width of the fingers collapses onto a master curve irrespective of the membrane thickness and increases linearly with $h$ as shown in Fig.~\ref{fig:geom_of_h}(a). The length of the fingers (Fig.~\ref{fig:geom_of_h}(b)) also increases approximately linearly with $h$, but the gradient increases with the membrane thickness. 
\begin{figure}
\centering
\includegraphics[width=1.0\linewidth]{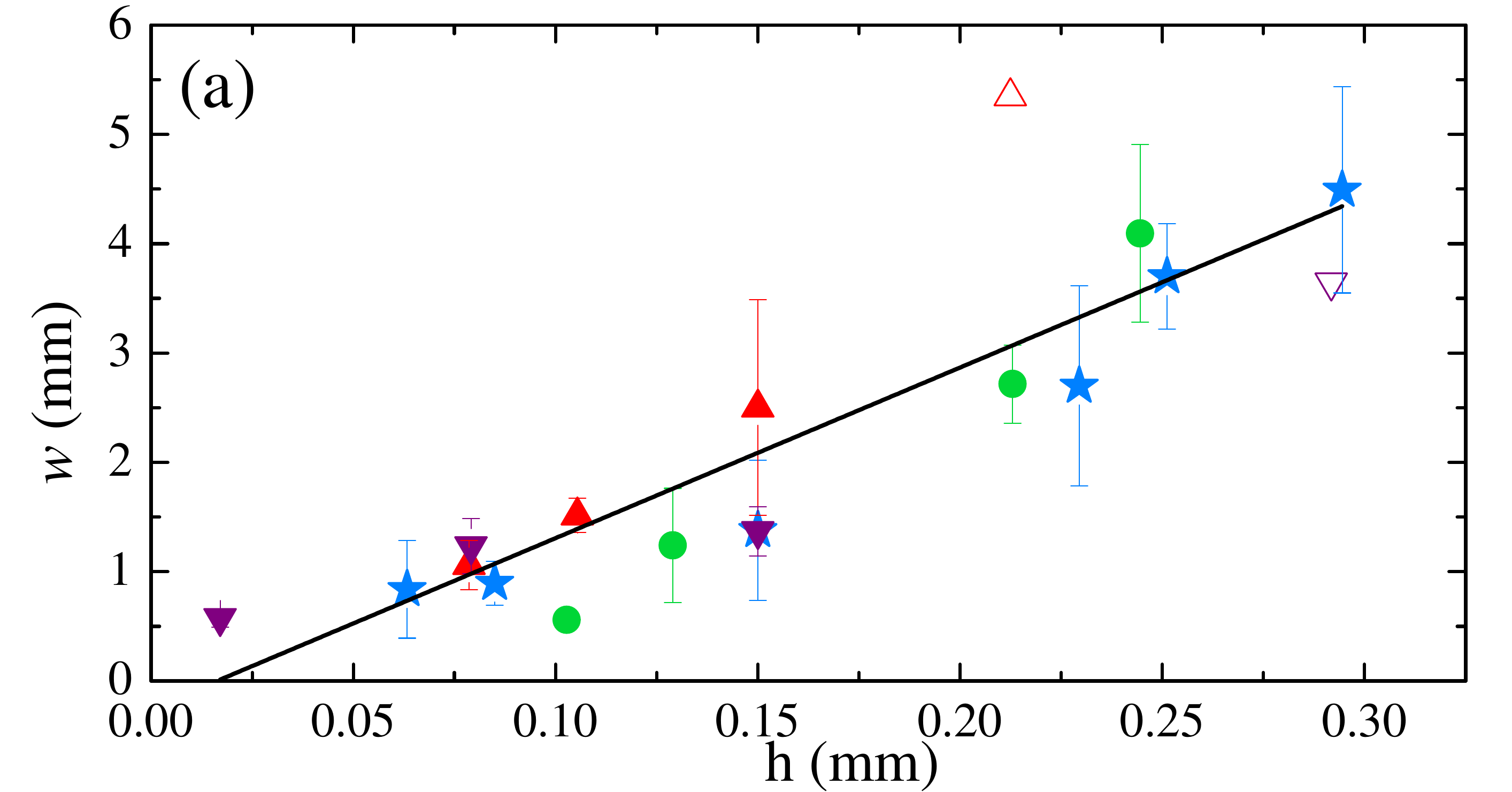}
\includegraphics[width=1.0\linewidth]{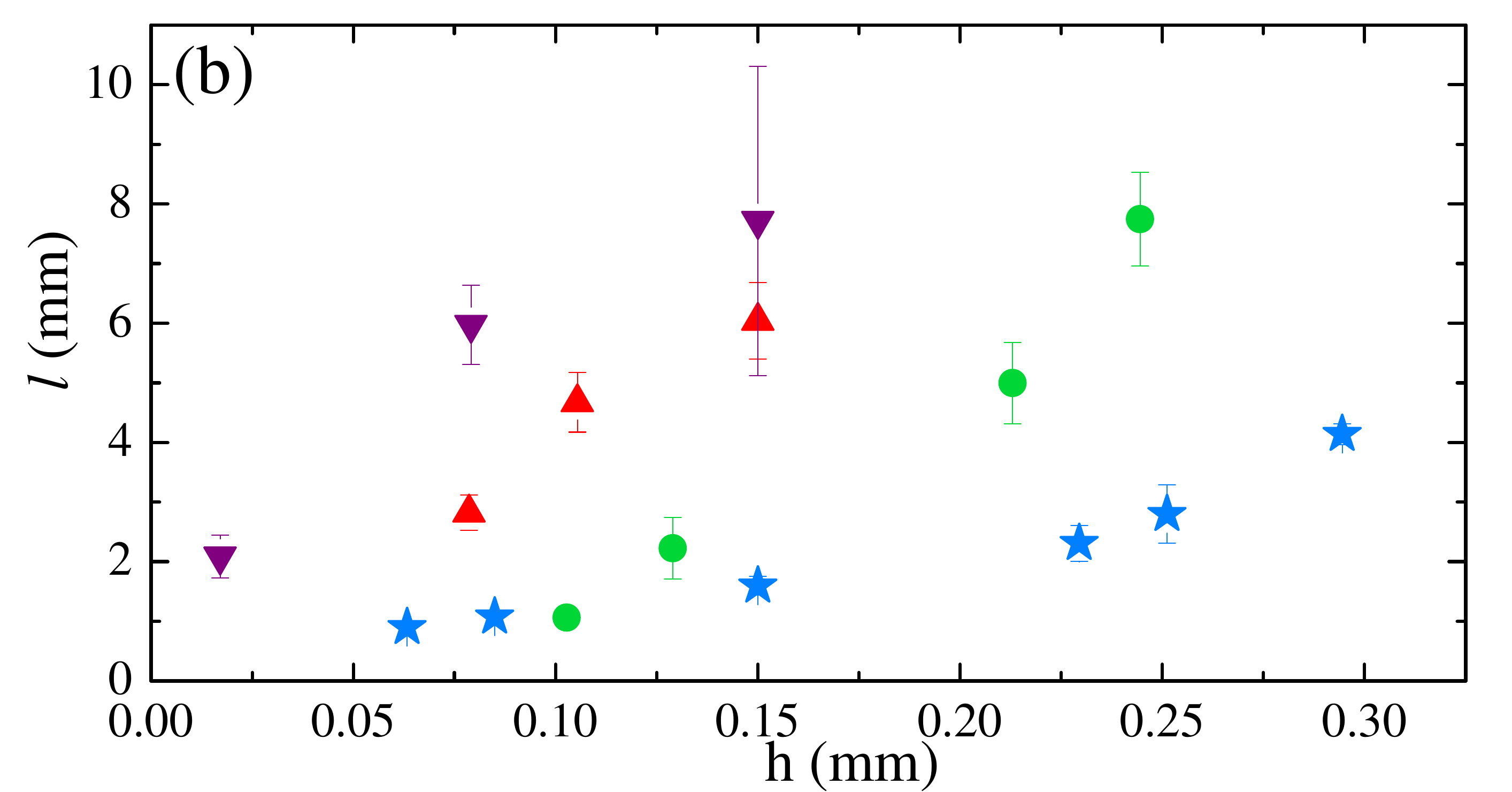}
\caption{Evolution of the finger width (a) and length (b) in the saturated regime with the thickness of the oil layer. $\mathrm{Ca}=0.88\pm0.03$ for $b=210$~$\mu$m, $\mathrm{Ca}=0.62\pm0.02$ for $b=340$~$\mu$m, $\mathrm{Ca}=1.38\pm0.05$ for $b=460$~$\mu$m and $\mathrm{Ca}=3.2\pm0.1$ for $b=570$~$\mu$m. The solid line is a linear lest-squares fit to all the data for $w$, excluding tip-splitting points (empty symbols). \label{fig:geom_of_h}}
\end{figure}
\paragraph*{}
The data obtained with the softest latex membrane are the closest to onset. The measured finger width scales as $1/\sqrt{\mathrm{Ca}}$, as shown in Fig.~\ref{fig:geom_of_Ca}(a). This scaling is consistent with the linear stability analysis for a flat interface propagating in a rigid rectangular channel, which predicts that the most unstable wavelength is proportional to $h/\sqrt{\mathrm{Ca}}$~\citep{saffman1958penetration}. This indicates that the average finger width is essentially set by the fluid mechanical instability. Moreover, $w$ remains proportional to $h$ in the saturated regime, indicating that the scaling at onset still holds in the saturated regime. A linear dependence of the wavelength on the layer depth is also  present in instabilities affecting thin layers of soft solids~\cite{chaudhury2015adhesion}, but we believe our results to be among the first to demonstrate this property in flowing viscous layers. By increasing the membrane thickness (and the transverse pre-tension) we increase $T$, which reduces the peeling angle (presented in the inset of Fig.~\ref{fig:length}). Larger values of the peeling angles $\theta$ have been shown to have a stabilizing effect on the interface, enhancing the contribution of restoring capillary forces and reducing that of destabilizing viscous forces, hence delaying the onset of fingering~\citep{pihler2012suppression, al2012control}. We never observed the total suppression of the instability using latex membranes in our compliant channel, because the critical Ca needed to propagate the uniform peeling mode is larger than that at the onset of finite-depth fingering. However, peeling of the thin silicone membrane occurred at larger values of $\theta$ for a given $\mathrm{Ca}$ and stabilized a uniformly flat interface. By consistently varying the peeling angle, we show that $\theta$ also has an effect on the fingering pattern: it does not affect the wavelength of the pattern, as shown by the collapse of all data in Fig.~\ref{fig:geom_of_h}(a), but steeper peeling fronts are more stable in a sense that they grow shorter fingers, while shallower peeling angles promote the growth of long fingers.
\begin{figure}
\centering
\includegraphics[width=1.0\linewidth]{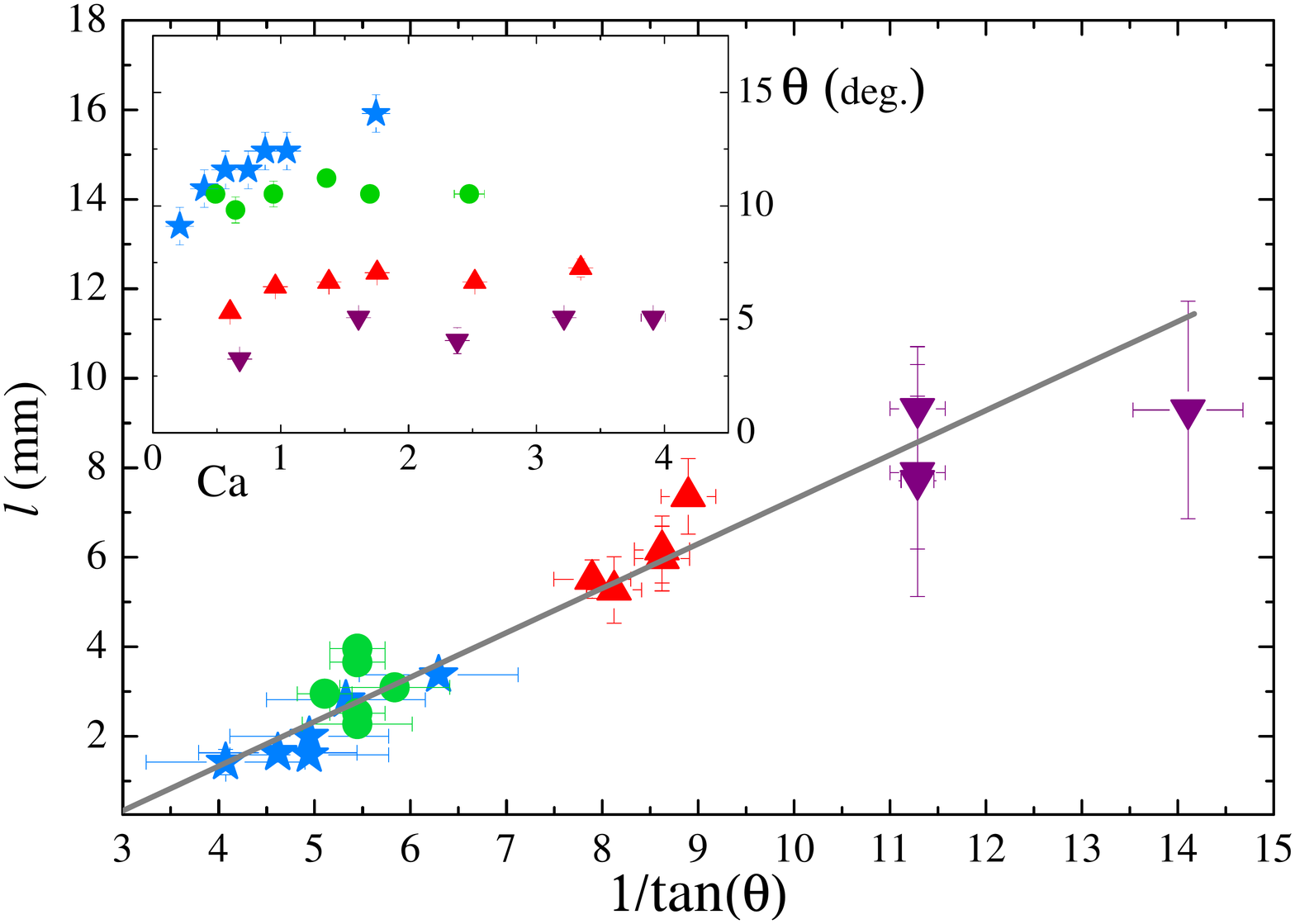}
\caption{Length of the fingers as a function of the peeling angle for $h=150$~$\mu$m. The red line is linear least-squares fit to the data. The inset gives the variation of the peeling angle with the capillary number for $h=150$~$\mu$m.\label{fig:length}}
\end{figure}
\paragraph*{}
To quantify the effect of the peeling angle on the finger length, we plot $l$ as a function of $1/\tan \theta$ in Fig.~\ref{fig:length} for all the data points shown in Fig.~\ref{fig:geom_of_Ca}(b). The length of the fingers is approximately proportional to $1/\tan \theta$, which indicates that they extend to a critical reopening height, which is around 1~mm in this case. We expect that this critical height depends on the volume of fluid available to form oil walls between the fingers, consistently with Fig.~\ref{fig:geom_of_h}(b), where the length of the fingers increases the liquid layer thickness $h$. This first quantification of the effect of the peeling angle on the pattern generated by viscous fingering opens up perspectives for controlled pattern formation: for a given $h$, the width of the pattern is set by the thickness of the liquid layer, while the length is selected by the peeling angle. From a fundamental point of view, spanning across a range of peeling angles allows us to explore the evolution of the fingering pattern as the channel transitions from compliant (steep peeling front) to rigid (limit of $\theta=0$) boundaries. In the rigid limit, channel expansion no longer restrains the axial extension of the fingers. The divergence of the finger length as $\theta$ tends to 0 is a first step towards recovering the classical Saffman-Taylor finger, which grows infinitely long in a rigid uniform channel. Although we could not observe such behavior, we would expect the constant-depth fingers to interact as they lengthen, as observed in viscous peeling experiments~\cite{mcewan1966peeling}. The study of elastic fingering allows us to make a connection between two classical viscous fingering problems, and in the light of the present results we conjecture that the printer's instability gives way to the classical Saffman-Taylor finger in a continuous transition as the stabilizing taper vanishes.
\section*{Acknowledgements}
This work was supported by the Leverhulme Trust under grant number RPG-2014-081.

\end{document}